\documentclass[letterpaper,10pt]{article} 
\usepackage{verbatim}
\usepackage{opticameet3}

\newcommand\authormark[1]{\textsuperscript{#1}}

\usepackage{amsmath,amssymb}
\usepackage[colorlinks=true,bookmarks=false,citecolor=blue,urlcolor=blue]{hyperref} %pdflatex

\begin{document}

\title{Defect-Induced Strain-Tunable Photoluminescence in AgScP$_2$S$_6$}

\author{Abhishek Mukherjee\authormark{1}, Damian Wlodarczyk\authormark{2},  Ajeesh K. Somakumar\authormark{2}, Piotr Sybilski\authormark{2}, Ryan Siebenaller\authormark{3,4}, Michael A. Susner\authormark{4}, Andrzej Suchocki\authormark{2} and Svetlana V. Boriskina\authormark{1,*}}

\address{\authormark{1} Massachusetts Institute of Technology, Cambridge, MA 02139, USA\\
\authormark{2} Institute of Physics, Polish Academy of Sciences, Warsaw, Poland\\
\authormark{3} Department of Materials Science and Engineering, The Ohio State University, Columbus, Ohio 43210, USA\\
\authormark{4} Materials and Manufacturing Directorate, Air Force Research Laboratory, Wright-Patterson AFB, Ohio 45433, USA}

\email{\authormark{*}sborisk@mit.edu}

\begin{abstract}
Metal thiophosphates (MTPs) are a large family of 2D materials that exhibit large structural and chemical diversity. They also show promise for applications in energy harvesting and photodetection. Strain and defect engineering have previously been demonstrated as useful mechanisms to tune several properties of MTPs such as resistivity, magnetic state, and electronic band gap. However, the effect of these stimuli on engineering tunable light emission in MTPs remains unexplored. Here, we show experimentally that structural defects in metal thiophosphate AgScP$_2$S$_6$ are prominent in exhibiting photoluminescence, which is likely driven by the defect-state-to-conduction-band transitions and can be further tuned by temperature-induced strain gradients. 
\end{abstract}

\section{Introduction}
Over the course of the last few decades, 2D van-der Waals layered materials have become a hotspot for material scientists, giving rise to the discovery of graphene, transition-metal dichalcogenides (TMDCs), and black phosphorus.  Recently, a family of layered metal thio(seleno)phosphates (MTPs) have gathered attention owing to their unique magnetic properties \cite{du2016weak,li2019intrinsic}, applications in energy harvesting \cite{wang2017two,mayorga2017layered,tiwari2020crystal,takeuchi2018sodium,fan2018high,yin2019cu,tanimoto2020enargite}, and interesting nonlinear optical effects \cite{kang2015metal,liang2017mid}. While hexagonal boron nitride (hBN) is exceptional for its strong, thickness-dependent optical nonlinearities \cite{kim2013stacking,li2013probing}, novel photonic avenues\cite{caldwell2019photonics,moon2023hexagonal,ogawa2023hexagonal,kumbhakar2021advance}, and interesting optical and electronic properties \cite{caldwell2019photonics,yi2019optical,moon2023hexagonal,shaik2021optical,kumbhakar2021advance}, it is unsuitable for low-energy operations due to its wide band gap ($\sim$ 6 eV). Similarly, transition metal dichalcogenides (TMDs) have enabled exciting electronic \cite{jiang2022flexible,wang2012electronics,gong2017electronic}, optoelectronic \cite{gong2017electronic,thakar2020optoelectronic,jing2020tunable,long2019progress,wang2012electronics,zhao2022advances} and photonic applications \cite{thakar2020optoelectronic,xia2014two,mak2016photonics,wang2019recent}, but their small bandgaps ($\sim$ 1-2 eV) limit their use in high-energy devices.
MTPs complete the multi-spectral library of 2D photonic materials as they possess with bandgaps between 1.6 and 4 eV, which provide the opportunity to fill the void between narrow band gap materials such as TMDs and wide band gap hBN. Physical properties of MTPs can be tuned via strain engineering, resulting in demonstrations of superconductivity \cite{he2023pressure,wang2018emergent} and insulator-to-metal transitions \cite{matsuoka2021pressure,matsuoka2023pressure,susilo2020band}, magnetic state tuning \cite{chittari2016electronic,coak2021emergent,ni2021imaging,zhang2021strain}, and optoelectronics \cite{fang2023pressure,han2021high}. TMDs can also form interesting hetero-structures with well-known TMDCs \cite{li2023proximity}, which further expands their range of functionalities. 

However, light emission properties of MTPs are much less explored, especially under symmetry-breaking strain deformations. This research direction is very promising since strain and defect engineering have been proven to be powerful and versatile mechanisms to tailor photonic properties of various material families, unlocking new functionalities in light harvesting and generation. For instance, Zhong et al \cite{zhong2023intercalation} showed that defect engineering can play a significant role in enhancing the electrochemical performance of MnPS$_3$. In this work, we explore the possibility of defect engineering for enhanced visible light emission in AgScP$_2$S$_6$, a new, recently synthesized member of the MTP family.

AgScP$_2$S$_6$ is a 2D direct bandgap semiconductor, which exhibits an extraordinarily large two-photon absorption coefficient predicted to be associated with defect/virtual states \cite{mushtaq2023agscp2s6}. Inspired by this prediction, we investigate photoluminescence (PL) in AgScP$_2$S$_6$ as an excellent probe for defects and explore the opportunities to achieve tunable light emission from MTPs via strain engineering. In this work, we show experimentally that (i) defects play a critical role in PL emission from AgScP$_2$S$_6$ bulk crystals and (ii) defect-mediated PL can be further tuned by temperature-induced strain gradients. Temperature-dependent absorption and Raman measurements have been used to characterize the defects and distinguish them from the bulk of the material.

\begin{figure}[h!]
    \centering
    \includegraphics[width=1\linewidth]{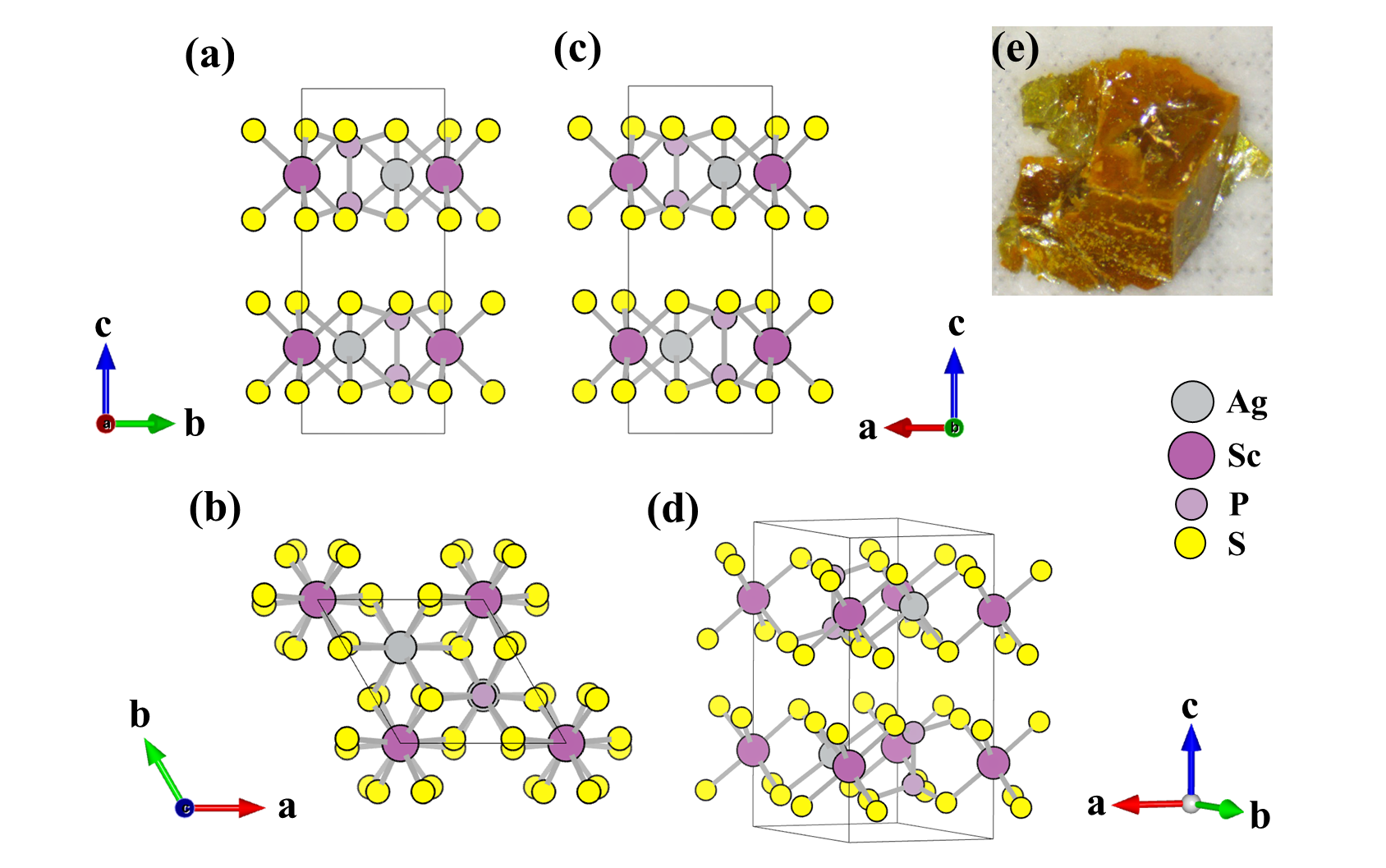}
    \caption{ AgScP$_2$S$_6$ crystal lattice and visual appearance. (a)-(d) Schematics of the AgScP$_2$S$_6$ crystal lattice under different axial configurations. (e) A photograph of an AgScP$_2$S$_6$ bulk crystal.}
    \label{fig:schematic}
\end{figure}

\section{Results}
Figures \ref{fig:schematic}a-d introduce AgScP$_2$S$_6$ material by presenting structural schematics of its crystal lattice, which belongs to the $P\overline{3}1\textit{c}$ symmetry group. The defining common feature for compounds in the MTP family is the [P$_2$S$_6$]$^{4-}$ anion sublattice located within an individual lamella of the layered crystal. This specific compound is of the type M$^{1+}$M$^{3+}$P$_2$S$_6$ \cite{mushtaq2023agscp2s6}.
The sulfur atoms make up almost closed packed surfaces of each layer. From Figs. \ref{fig:schematic}b,d, it can be seen that the spatial arrangement of atoms leads to octahedrally coordinated sites formed in/within the lattice, which are filled by Ag and Sc ions, or P--P dimers. Fig. \ref{fig:schematic}e shows an image of the bulk crystal, which has been synthesized via a vapor transport technique \cite{susner2017metal}, see the Methods section for detail. 

\begin{figure}
    \centering
    \includegraphics[width=1\linewidth]{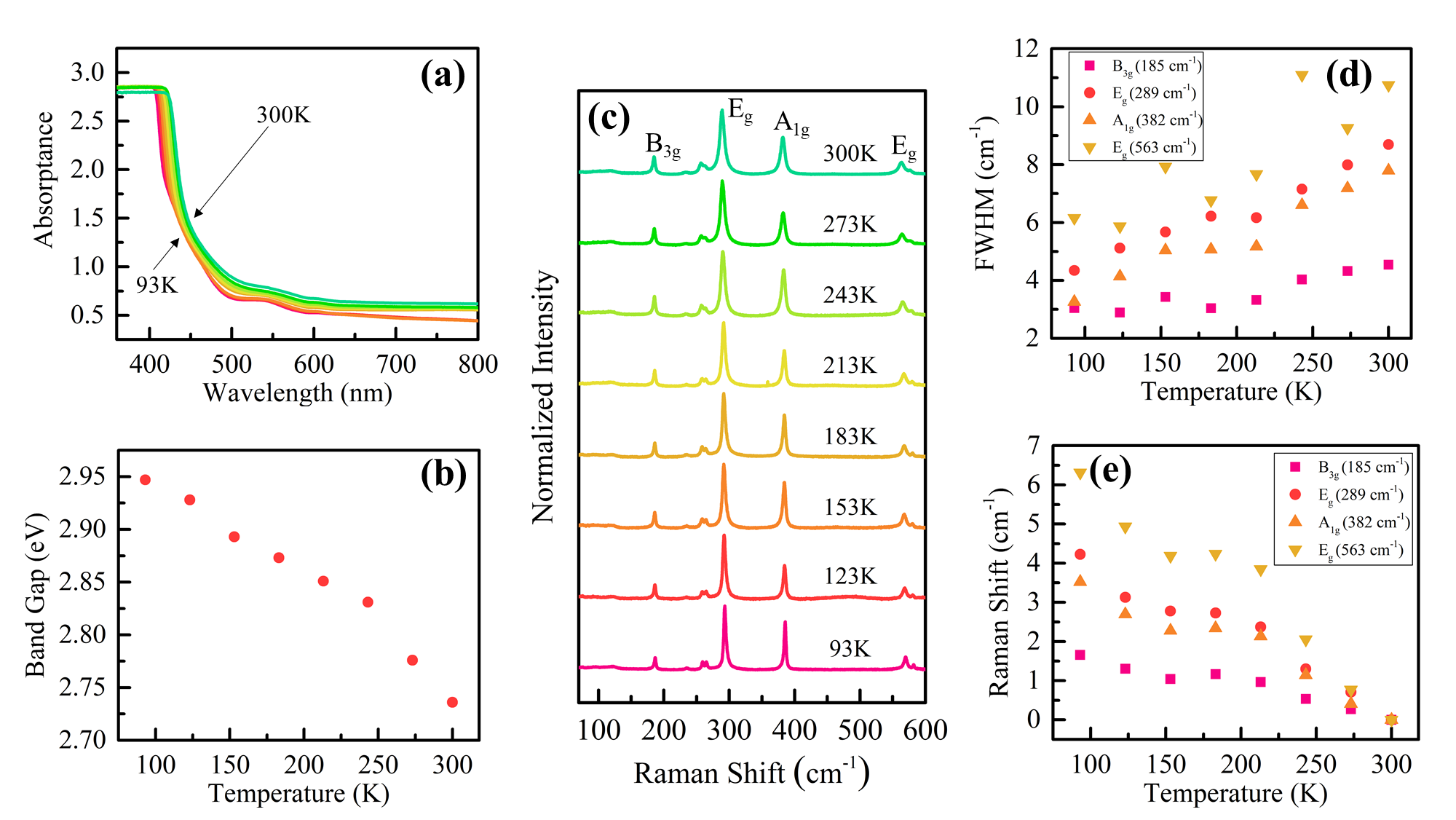}
    \caption{Optical and vibrational properties of AgScP$_2$S$_6$. (a) Absorption spectra of AgScP$_2$S$_6$ measured as a function of temperature under ambient pressure. (b) Band gap width values extracted from the spectra in panel (a). (c) Temperature-dependent Raman spectra measured from 300K to 93K. Four prominent Raman modes are labeled in the spectra taken at 300K. (d) Full-width at half maximum (FWHM) in cm$^{-1}$ as a function of temperature for the Raman peaks shown in (c). (e) Relative shifts in Raman peaks from their corresponding spectral positions measured at 300K.}
    \label{fig:abs}
\end{figure}

Figure \ref{fig:abs}a illustrates the measured UV-Visible absorption spectrum for AgScP$_2$S$_6$ as a function of temperature. At room temperature (300K), the absorption was found to decrease rapidly at wavelengths larger than $\sim$420 nm. This drop is characteristic of the band edge absorption typically observed in direct band gap semiconductors. The same trend is witnessed as temperature is decreased from 300K down to 93K. In Fig. \ref{fig:abs}b, we plot the estimated band gap energy values as a function of temperature, which are extracted from the absorption spectra by using a standard fitting procedure \cite{ghobadi2013band}. The band gap value obtained at ambient conditions was found to be 2.74 eV. This is consistent with the previous measurement reported in the literature \cite{mushtaq2023agscp2s6}. From Fig. \ref{fig:abs}b, it can be seen that the band gap energy increases almost linearly with decreasing temperature.

Figure \ref{fig:abs}c shows Raman spectra of the material spanning the same temperature range. Typically in metal thiophosphates, a band below 150 cm$^{-1}$ is attributed to cation vibrations\cite{senkine1989raman,balkanski1987effects,scagliotti1987raman}. We also identify a peak at 185 cm$^{-1}$, most likely due to the B$_{3g}$ mode of vibration \cite{gjikaj2006rb2p2s6}, and the rotation of the PS$_3$ group \cite{oliveira20232d}. We identify phonon modes originating from P$_2$S$_6$ units with D$_{3d}$ symmetry (center of inversion along PS$_3$ groups) as A$_{1g}$ and E$_g$ \cite{gusmao2017role,hangyo1988raman,long2017isolation,silipigni2005x}. Four of such prominent Raman modes are marked in the spectrum measured at 300K (Fig. \ref{fig:abs}c). (Note that we did not perform any calculation of our own to mark these modes, they have been identified by comparing data from previously published literature on metal thiophosphates). As the temperature is decreased, we observe changes in peak positions and full widths at half maximum (FWHM). We also find that the material exhibits no phase transitions at lower temperatures. The changes in Raman peak positions and FWHM are calculated with reference to the values measured at 300K and are shown in Figs. \ref{fig:abs}d and \ref{fig:abs}e, respectively. The temperature coefficients are extracted from the Raman data and presented in Table \ref{tab:my_label}. 

From Fig. \ref{fig:abs}d, we see that phonon modes move to higher energies as temperature is reduced, which can be explained by tighter chemical bonding in the compound leading to higher vibrational energies. In turn, the FWHM of the peaks can be related to the phonon lifetime $\tau$ using the Heisenberg Uncertainty relation:

\begin{equation}
    \frac{\Delta E}{\hslash} = \frac{1}{\tau},
\end{equation}

where $\Delta E$ is the Raman FWHM in units of cm$^{-1}$, and $\hslash$ is the Planck's constant. We identify two main contributions to the phonon lifetime: (i) phonon-phonon scattering -- which is the characteristic anharmonic decay of a phonon into two or more phonons, and (ii) phonon-carrier scattering -- caused by the perturbation in the translational symmetry of the crystal due to the presence of defects. It is difficult to separate the relative contributions of these mechanisms to the overall lifetime and further studies would be required to investigate their individual contributions. From Fig. \ref{fig:abs}b, it can be seen that FWHM linearly decreases as the temperature is decreased. From equation (1), it is apparent that the phonon lifetime decreases with increasing temperature. The decrease occurs due to increased thermal occupancy and interaction of phonons at higher temperatures \cite{beechem2008temperature}, which in turn increases the rate of phonon-phonon and carrier-phonon scattering. \\

\begin{table}[]
    \centering
    \begin{tabular}{||c c c c||}
     \hline
    Raman Mode & Peak Position at 300K (cm$^{-1}$) & dE/dT (cm$^{-1}$/K) & d(FWHM)/dT (cm$^{-1}$/K) \\
    \hline\hline
        B$_{3g}$ & 185.0 & -0.007 ± 0.001 & 0.009 $\pm$ 0.001\\
        E$_g$ & 289.2 & -0.019 $\pm$ 0.002 & 0.020 $\pm$ 0.001    \\
        A$_{1g}$ & 381.9 & -0.016 $\pm$ 0.002 & 0.021 $\pm$ 0.001\\
        E$_g$ & 563.4 & -0.029 $\pm$ 0.003 & 0.018 $\pm$ 0.005\\
      \hline  
    \end{tabular}
    \caption{Temperature coefficients of AgScP$_2$S$_6$ vibrational modes determined from Raman Spectroscopy.}
    \label{tab:my_label}
\end{table}

\begin{figure}[h!]
    \centering
    \includegraphics[width=1\linewidth]{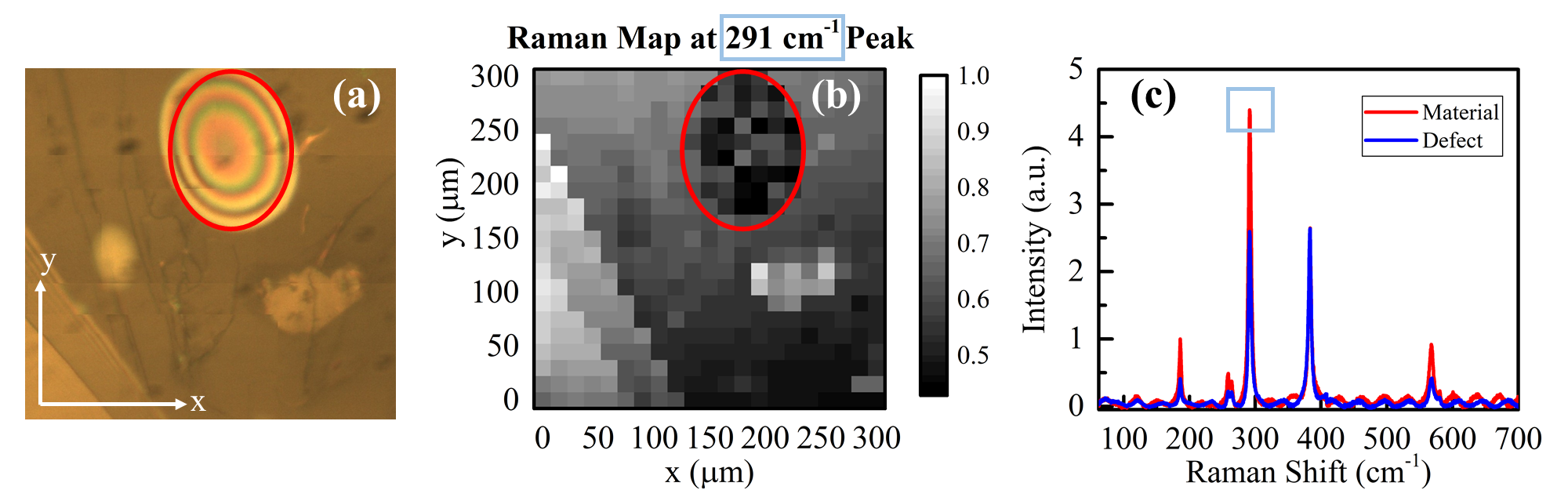}
    \caption{Raman signals collected from structural defects reveal defect-induced photoluminescence. (a) Example of a structural `peacock-like' defect region (area marked by the red circle). (b) Raman map of the spatial area shown in the image in (a) plotted for the Raman peak at 289 cm$^{-1}$ marked by the blue square in (c). (c) Raman signature for a structural defect (blue), compared with the bulk (red).}
    \label{fig:defect}
\end{figure}

Figure \ref{fig:defect} shows the defects in bulk crystals of AgScP$_2$S$_6$ (Fig. \ref{fig:defect}a) and illustrates the difference in the Raman spectra (Fig. \ref{fig:defect}b,c) collected from the bulk (red) and defect regions (blue). These microscopic defects are introduced during the process of material growth. We find that most of the observed defects are structural (physical), i.e. they have the same overall Raman spectrum signature as the rest of the material (compare the normalized spectra in Fig. \ref{fig:defect}c). 

However, the Raman signal from the `defect region' within the red circle in Figs. 3a-b, plotted prior to background subtraction, is weak. This local Raman signal attenuation may be caused by the Raman scattering process interference with the material photoluminescence under the laser pump excitation. PL interference is one of the common underlying mechanisms contributing to the Raman lines suppression and may play such a role in AgScP$_2$S$_6$, as this material exhibits a PL spectrum overlapping with the Raman spectrum. Using a hypothesis that the inverse correlation between Raman and PL signals can help to identify the comparatively more photo-luminescent areas on the sample, we find and map these areas via Raman imaging spectroscopy. 

An example of a Raman map taken around a structural defect region is shown in Fig. 3b. The mapping is done at the strongest Raman peak at $\sim$ 291 cm$^{-1}$ marked by a light blue square in Fig. 3c. The area shown in the red circle in Fig. 3b exhibits the weakest Raman intensity, which spatially overlaps with the structural defect at this location, as shown in Fig. 3a.

\begin{figure}[h!]
    \centering
    \includegraphics[width=1\linewidth]{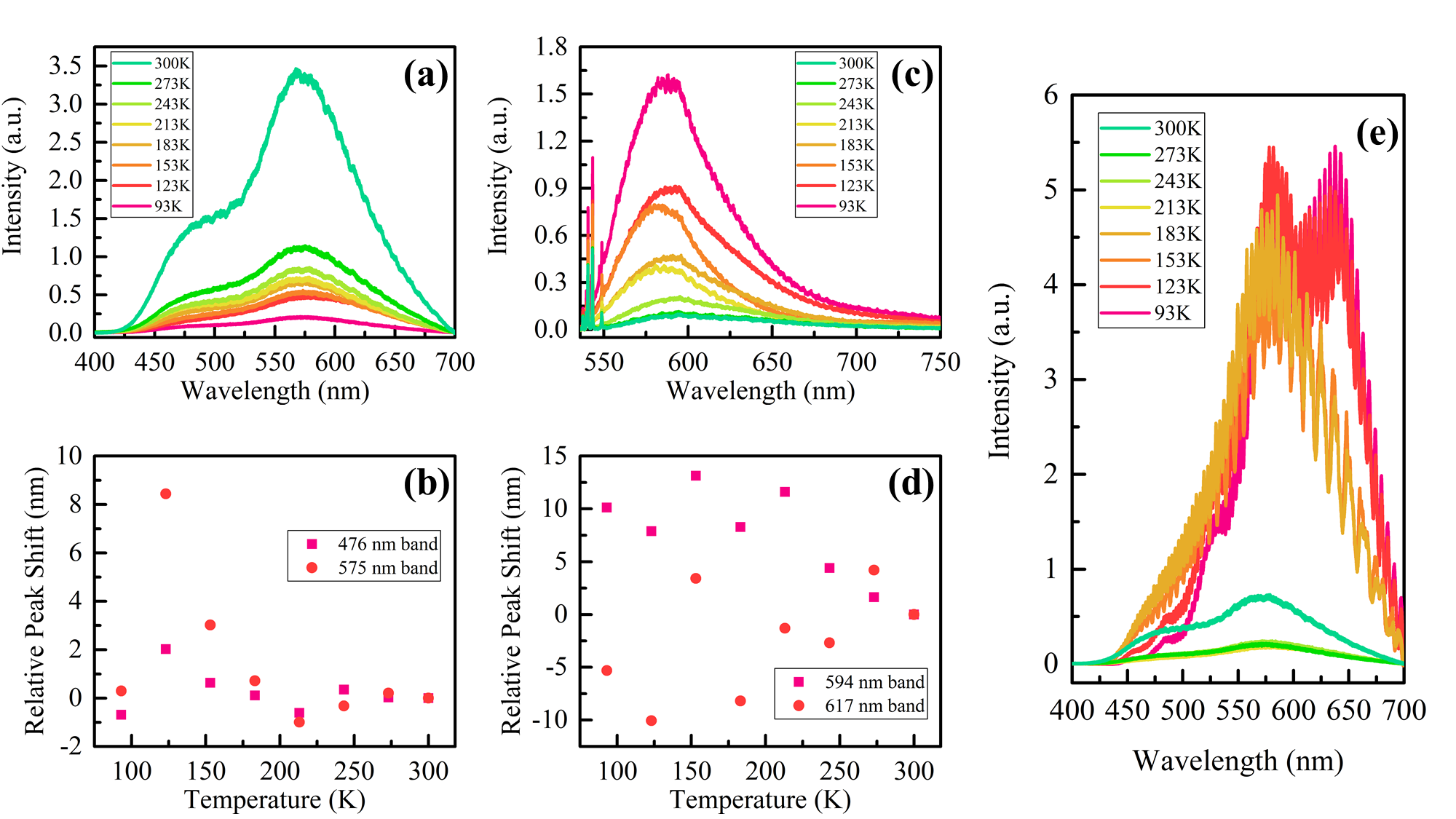}
    \caption{Structural defects create local PL centers. PL intensity as a function of temperature for (a) the bulk material, (e) a structural defect under 325 nm laser excitation. (c) PL intensity as a function of temperature for the bulk material under 532 nm excitation. Peak positions of the PL spectral lines corresponding to (a), (c) are shown in (b), (d) respectively. }
    \label{fig:PL}
\end{figure}

 To test our hypothesis that these Raman-weak structural defects exhibit strong PL emission as compared to the rest of the material, we correlate our findings with the PL spectra collected from both the bulk region and the defect region, which are measured as a function of temperature. PL measurements have been performed under three distinct operational scenarios -- (i) at a laser excitation energy higher than the band gap (325 nm, 3.81 eV $> E_g =$ 2.74 eV), (ii) at an excitation within the band gap (532 nm, 2.33 eV $< E_g =$ 2.74 eV), and (iii) PL spectra obtained by directly exciting a structural defect (of $\sim$ 50 $\mu$m in diameter) with the 325 nm laser line having $\sim 10 \mu$m diameter laser spot. The results are shown and compared in Fig. \ref{fig:PL}.

Figure \ref{fig:PL}a reveals PL quenching with exponential decrease in intensity as the temperature is decreased from 300K to 93K. This trend is explained by the evolution of the bandgap size with temperature, which has been extracted from the temperature-dependent absorption measurements and plotted in Fig. \ref{fig:abs}b. The bandgap widens with decreasing temperature, decreasing the probability of interband transitions and leading to the decrease in the bulk PL signal observed in Fig. \ref{fig:PL}a . The PL spectra can be deconvoluted into two bands peaking at 476 nm and 575 nm approximately. The behavior of the peak positions for these bands, relative to the peak position measured at 300K, are shown in Fig. \ref{fig:PL}b. Both luminescent bands follow a similar trend -- red-shifting on average at lower temperatures. 

Figure \ref{fig:PL}c summarizes the PL spectra collected in the case when we perform the same PL spectral experiment with the 532 nm laser line. Here, we see a completely opposite trend from than observed in Fig. \ref{fig:PL}a, with PL emission intensity exponentially increasing with decreasing temperature. In the next paragraph, we propose a rationale for this finding. Again, we can deconvolute the PL spectra in Fig. \ref{fig:PL}c to get two luminescent bands peaking at roughly 594 nm and 617 nm. We track the relative peak positions of these bands in Fig. \ref{fig:PL}d. It can be seen that the two bands move further apart with decreasing temperature, with the 594 nm band red-shifting and the 617 nm band blue-shifting.

Case (iii) is shown in Fig. \ref{fig:PL}e, where we excite a structural defect of $\sim$ 50 $\mu$m in diameter with the laser spot of $\sim 10 \mu$m in diameter. PL emission from the defect is quenched as the temperature decreases to 213K, consistent with the trend we expect for a 325 nm laser excitation. However, in the temperature range 183-93K, we observe a substantial increase in PL intensities. This supports our hypothesis of defect states lying within the band gap, which make possible defect-to-band transitions excited by the 325 nm laser pump. The increase in the PL signal is accompanied by interference effects manifested as Fabry-Perot fringes, which can be attributed to the structural defect acting as a resonant cavity. We hypothesize that the increase in PL can be either due to unexplained defect-carrier statistics or due to the temperature-induced `wrinkling' of the defect creating strain gradients, which modify both the defect geometry and the electron band structure. The latter hypothesis is supported by the emergence of new photoluminescent bands (forming a prominent double-peak feature) seen in the PL spectra collected at 123K and 93K.

\section{Experimental Section}
\label{sec:headings}

\subsection{Material Synthesis}

We synthesized AgScP$_2$S$_6$ via conventional vapor transport methods \cite{susner2017metal}. Briefly, we combined Ag foil (Johnson Matthey, 99.99\%), Sc metal (Alfa Aesar pellet, 99.999\%, rolled thin) P chunks (Alfa Aesar Puratronic, 99.999+\%), and S pieces (Alfa Aesar Puratronic, 99.999\%) together in a near-stoichiometric ratio (we used 10\% excess P) in an evacuated quartz ampoule (2mm wall thickness, 22 mm OD, 10 cm in length) together with $\sim$100 mg I2 crystals (Alfa Aesar, 99.8\%). We placed the sealed ampoule in a single-zone tube furnace, heated to 750$^{\circ} $C over a period of 20 hours, held at that temperature for 100 hrs, and cooled over a period of 20 hrs. The resulting crystals were of maximum size of 5 mm $\times$ 5 mm in area and $\sim$200 $\mu$m in thickness. Sample compositions were determined by subjecting at least three distinct single-crystal specimens of each batch to multiple-spot scanning electron microscopy/energy-dispersive X-ray spectroscopy (SEM/EDS) analysis (9-12 spots total per batch) using a Thermo Scientific UltraDry EDS spectrometer joined with a JEOL JSM-6060 SEM. The compositions were stoichiometric, within error.

\subsection{Instruments and Methods}
For observing defect-assisted Photoluminescence (PL), the samples were excited with a near-UV (325 nm) Kimmon-Koha IK3201R-F model laser. PL data was collected using a Triax 320 monochromator from ISA Yobin Yvon-Spex equipped with a Spectrum One CCD detector. A Cary 5000 spectrophotometer was used to obtain absorption spectra in the UV-Visible range. Raman spectral data were acquired using a Monovista CRS+ system from S\&I Ltd, equipped with a 0.75 m long monochromator. The resolution was approximately  0.5 cm$^{-1}$. The setup was operated on Trivista Software. A 532 nm laser was used, set at a power density of 7.5 mW/$\mu$m$^2$. Raman, Absorption and PL data were collected at both room temperature and low temperatures under ambient pressure. Low temperature measurements were performed on Linkam FTIR600 table under steady N$_2$ flow to prevent air and moisture from freezing. Line positions and FWHM in the Raman spectra were established by fitting the observed spectra with Lorentzian curves. Gaussian curves were used to deconvolute the observed PL data and to measure peak positions.

\section{Conclusions and Outlook}
Structural defects have been characterized and shown to be the main contributor towards low-temperature PL emission in AgScP$_2$S$_6$. However, the origin of these defect states needs further studies. We plan to take further steps to characterize these defects using advanced techniques such as EPR (Electron Paramagnetic Resonance) and to systematically engineer the defects by using energetic electron beams as reviewed in \cite{jiang2019defect}. To explore the possibility of using strain gradients to create new PL bands and to increase emission intensity, we are developing a fabrication pipeline to create such gradients in MTP materials  using nano-pillars and gratings on a CMOS platform. This will be a crucial step towards our final goal of achieving strain-tunable photon sources by using materials from the MTP family.\\

\paragraph{Acknowledgements}
This work has been supported by the MIT-Poland Lockheed Martin Seed Fund and the ARO MURI Grant No. W911NF-19-1-0279. Abhishek Mukherjee appreciates the support provided by the Siebel Scholarship. Michael A. Susner acknowledges support of the Air Force Office of Scientific Research (AFOSR) Grant No. LRIR 23RXCOR003 and AOARD MOST Grant No. F4GGA21207H002 as well as general support from the Air Force Materials and Manufacturing (RX) and Aerospace Systems (RQ) Directorates. This work was also partially supported by the Polish National Science Center SHENG-2 Grant No. 2021/40/Q/ST5/00336 and Preludium Grant No. 2019/33/N/ST5/02317.

\paragraph{Conflict of Interest} The authors declare no conflict of interest.

\bibliographystyle{unsrt}
\bibliography{sample}

\begin{thebibliography}{10}

\bibitem{du2016weak}
Ke-zhao Du, Xing-zhi Wang, Yang Liu, Peng Hu, M~Iqbal~Bakti Utama, Chee~Kwan Gan, Qihua Xiong, and Christian Kloc.
\newblock Weak van der waals stacking, wide-range band gap, and raman study on ultrathin layers of metal phosphorus trichalcogenides.
\newblock {\em ACS nano}, 10(2):1738--1743, 2016.

\bibitem{li2019intrinsic}
Hui Li, Shuangchen Ruan, and Yu-Jia Zeng.
\newblock Intrinsic van der waals magnetic materials from bulk to the 2d limit: new frontiers of spintronics.
\newblock {\em Advanced Materials}, 31(27):1900065, 2019.

\bibitem{wang2017two}
Fengmei Wang, Tofik~Ahmed Shifa, Peng He, Zhongzhou Cheng, Junwei Chu, Yang Liu, Zhenxing Wang, Feng Wang, Yao Wen, Lirong Liang, et~al.
\newblock Two-dimensional metal phosphorus trisulfide nanosheet with solar hydrogen-evolving activity.
\newblock {\em Nano Energy}, 40:673--680, 2017.

\bibitem{mayorga2017layered}
Carmen~C Mayorga-Martinez, Zdenek Sofer, David Sedmidubsky, Stepan Huber, Alex Yong~Sheng Eng, and Martin Pumera.
\newblock Layered metal thiophosphite materials: magnetic, electrochemical, and electronic properties.
\newblock {\em ACS applied materials \& interfaces}, 9(14):12563--12573, 2017.

\bibitem{tiwari2020crystal}
Devendra Tiwari, Dominic Alibhai, David Cherns, and David~J Fermin.
\newblock Crystal and electronic structure of bismuth thiophosphate, bips4: an earth-abundant solar absorber.
\newblock {\em Chemistry of Materials}, 32(3):1235--1242, 2020.

\bibitem{takeuchi2018sodium}
Shigeo Takeuchi, Kota Suzuki, Masaaki Hirayama, and Ryoji Kanno.
\newblock Sodium superionic conduction in tetragonal na3ps4.
\newblock {\em Journal of Solid State Chemistry}, 265:353--358, 2018.

\bibitem{fan2018high}
Xiulin Fan, Jie Yue, Fudong Han, Ji~Chen, Tao Deng, Xiuquan Zhou, Singyuk Hou, and Chunsheng Wang.
\newblock High-performance all-solid-state na--s battery enabled by casting--annealing technology.
\newblock {\em ACS nano}, 12(4):3360--3368, 2018.

\bibitem{yin2019cu}
Xinxing Yin, Scott~A McClary, Zhaoning Song, Dewei Zhao, Brian Graeser, Changlei Wang, Niraj Shrestha, Xiaoming Wang, Cong Chen, Chongwen Li, et~al.
\newblock A cu 3 ps 4 nanoparticle hole selective layer for efficient inverted perovskite solar cells.
\newblock {\em Journal of Materials Chemistry A}, 7(9):4604--4610, 2019.

\bibitem{tanimoto2020enargite}
Takuya Tanimoto, Koichiro Suekuni, Taiki Tanishita, Hidetomo Usui, Terumasa Tadano, Taiga Kamei, Hikaru Saito, Hirotaka Nishiate, Chul~Ho Lee, Kazuhiko Kuroki, et~al.
\newblock Enargite cu3ps4: A cu--s-based thermoelectric material with a wurtzite-derivative structure.
\newblock {\em Advanced Functional Materials}, 30(22):2000973, 2020.

\bibitem{kang2015metal}
Lei Kang, Molin Zhou, Jiyong Yao, Zheshuai Lin, Yicheng Wu, and Chuangtian Chen.
\newblock Metal thiophosphates with good mid-infrared nonlinear optical performances: a first-principles prediction and analysis.
\newblock {\em Journal of the American Chemical Society}, 137(40):13049--13059, 2015.

\bibitem{liang2017mid}
Fei Liang, Lei Kang, Zheshuai Lin, and Yicheng Wu.
\newblock Mid-infrared nonlinear optical materials based on metal chalcogenides: structure--property relationship.
\newblock {\em Crystal Growth \& Design}, 17(4):2254--2289, 2017.

\bibitem{kim2013stacking}
Cheol-Joo Kim, Lola Brown, Matt~W Graham, Robert Hovden, Robin~W Havener, Paul~L McEuen, David~A Muller, and Jiwoong Park.
\newblock Stacking order dependent second harmonic generation and topological defects in h-bn bilayers.
\newblock {\em Nano letters}, 13(11):5660--5665, 2013.

\bibitem{li2013probing}
Yilei Li, Yi~Rao, Kin~Fai Mak, Yumeng You, Shuyuan Wang, Cory~R Dean, and Tony~F Heinz.
\newblock Probing symmetry properties of few-layer mos2 and h-bn by optical second-harmonic generation.
\newblock {\em Nano letters}, 13(7):3329--3333, 2013.

\bibitem{caldwell2019photonics}
Joshua~D Caldwell, Igor Aharonovich, Guillaume Cassabois, James~H Edgar, Bernard Gil, and DN~Basov.
\newblock Photonics with hexagonal boron nitride.
\newblock {\em Nature Reviews Materials}, 4(8):552--567, 2019.

\bibitem{moon2023hexagonal}
Seokho Moon, Jiye Kim, Jeonghyeon Park, Semi Im, Jawon Kim, Inyong Hwang, and Jong~Kyu Kim.
\newblock Hexagonal boron nitride for next-generation photonics and electronics.
\newblock {\em Advanced Materials}, 35(4):2204161, 2023.

\bibitem{ogawa2023hexagonal}
Shinpei Ogawa, Shoichiro Fukushima, and Masaaki Shimatani.
\newblock Hexagonal boron nitride for photonic device applications: A review.
\newblock {\em Materials}, 16(5):2005, 2023.

\bibitem{kumbhakar2021advance}
Partha Kumbhakar, Chinmayee Chowde~Gowda, and Chandra~Sekhar Tiwary.
\newblock Advance optical properties and emerging applications of 2d materials.
\newblock {\em Frontiers in Materials}, 8:721514, 2021.

\bibitem{yi2019optical}
Ya~Yi, Zhengbo Sun, Jia Li, Paul~K Chu, and Xue-Feng Yu.
\newblock Optical and optoelectronic properties of black phosphorus and recent photonic and optoelectronic applications.
\newblock {\em Small Methods}, 3(10):1900165, 2019.

\bibitem{shaik2021optical}
Akbar Basha Dhu-al-jalali-wal-ikram Shaik and Penchalaiah Palla.
\newblock Optical quantum technologies with hexagonal boron nitride single photon sources.
\newblock {\em Scientific reports}, 11(1):12285, 2021.

\bibitem{jiang2022flexible}
Dongting Jiang, Zhiyuan Liu, Zhe Xiao, Zhengfang Qian, Yiling Sun, Zhiyuan Zeng, and Renheng Wang.
\newblock Flexible electronics based on 2d transition metal dichalcogenides.
\newblock {\em Journal of Materials Chemistry A}, 10(1):89--121, 2022.

\bibitem{wang2012electronics}
Qing~Hua Wang, Kourosh Kalantar-Zadeh, Andras Kis, Jonathan~N Coleman, and Michael~S Strano.
\newblock Electronics and optoelectronics of two-dimensional transition metal dichalcogenides.
\newblock {\em Nature nanotechnology}, 7(11):699--712, 2012.

\bibitem{gong2017electronic}
Chuanhui Gong, Yuxi Zhang, Wei Chen, Junwei Chu, Tianyu Lei, Junru Pu, Liping Dai, Chunyang Wu, Yuhua Cheng, Tianyou Zhai, et~al.
\newblock Electronic and optoelectronic applications based on 2d novel anisotropic transition metal dichalcogenides.
\newblock {\em Advanced Science}, 4(12):1700231, 2017.

\bibitem{thakar2020optoelectronic}
Kartikey Thakar and Saurabh Lodha.
\newblock Optoelectronic and photonic devices based on transition metal dichalcogenides.
\newblock {\em Materials Research Express}, 7(1):014002, 2020.

\bibitem{jing2020tunable}
Yumei Jing, Baoze Liu, Xukun Zhu, Fangping Ouyang, Jian Sun, and Yu~Zhou.
\newblock Tunable electronic structure of two-dimensional transition metal chalcogenides for optoelectronic applications.
\newblock {\em Nanophotonics}, 9(7):1675--1694, 2020.

\bibitem{long2019progress}
Mingsheng Long, Peng Wang, Hehai Fang, and Weida Hu.
\newblock Progress, challenges, and opportunities for 2d material based photodetectors.
\newblock {\em Advanced Functional Materials}, 29(19):1803807, 2019.

\bibitem{zhao2022advances}
Mingyue Zhao, Yurui Hao, Chen Zhang, Rongli Zhai, Benqing Liu, Wencheng Liu, Cong Wang, Syed Hassan~Mujtaba Jafri, Aamir Razaq, Raffaello Papadakis, et~al.
\newblock Advances in two-dimensional materials for optoelectronics applications.
\newblock {\em Crystals}, 12(8):1087, 2022.

\bibitem{xia2014two}
Fengnian Xia, Han Wang, Di~Xiao, Madan Dubey, and Ashwin Ramasubramaniam.
\newblock Two-dimensional material nanophotonics.
\newblock {\em Nature Photonics}, 8(12):899--907, 2014.

\bibitem{mak2016photonics}
Kin~Fai Mak and Jie Shan.
\newblock Photonics and optoelectronics of 2d semiconductor transition metal dichalcogenides.
\newblock {\em Nature Photonics}, 10(4):216--226, 2016.

\bibitem{wang2019recent}
Xiaoting Wang, Yu~Cui, Tao Li, Ming Lei, Jingbo Li, and Zhongming Wei.
\newblock Recent advances in the functional 2d photonic and optoelectronic devices.
\newblock {\em Advanced Optical Materials}, 7(3):1801274, 2019.

\bibitem{he2023pressure}
Xinyi He, Zhitao Zhang, Zhenyu Ding, Shuyang Wang, Chunhua Chen, Xuliang Chen, Yonghui Zhou, Chao An, Min Zhang, Ying Zhou, et~al.
\newblock Pressure-induced superconductivity in the metal thiophosphate pb 2 p 2 s 6.
\newblock {\em Physical Review Materials}, 7(5):054801, 2023.

\bibitem{wang2018emergent}
Yonggang Wang, Jianjun Ying, Zhengyang Zhou, Junliang Sun, Ting Wen, Yannan Zhou, Nana Li, Qian Zhang, Fei Han, Yuming Xiao, et~al.
\newblock Emergent superconductivity in an iron-based honeycomb lattice initiated by pressure-driven spin-crossover.
\newblock {\em Nature Communications}, 9(1):1914, 2018.

\bibitem{matsuoka2021pressure}
Takahiro Matsuoka, Amanda Haglund, Rui Xue, Jesse~S Smith, Maik Lang, Antonio~M dos Santos, and David Mandrus.
\newblock Pressure-induced insulator--metal transition in two-dimensional mott insulator nips3.
\newblock {\em Journal of the Physical Society of Japan}, 90(12):124706, 2021.

\bibitem{matsuoka2023pressure}
Takahiro Matsuoka, Rahul Rao, Michael~A Susner, Benjamin~S Conner, Dongzhou Zhang, and David Mandrus.
\newblock Pressure-induced insulator-to-metal transition in the van der waals compound cops 3.
\newblock {\em Physical Review B}, 107(16):165125, 2023.

\bibitem{susilo2020band}
Resta~A Susilo, Bo~Gyu Jang, Jiajia Feng, Qianheng Du, Zhipeng Yan, Hongliang Dong, Mingzhi Yuan, Cedomir Petrovic, Ji~Hoon Shim, Duck~Young Kim, et~al.
\newblock Band gap crossover and insulator--metal transition in the compressed layered crps4.
\newblock {\em npj Quantum Materials}, 5(1):58, 2020.

\bibitem{chittari2016electronic}
Bheema~Lingam Chittari, Youngju Park, Dongkyu Lee, Moonsup Han, Allan~H MacDonald, Euyheon Hwang, and Jeil Jung.
\newblock Electronic and magnetic properties of single-layer m p x 3 metal phosphorous trichalcogenides.
\newblock {\em Physical Review B}, 94(18):184428, 2016.

\bibitem{coak2021emergent}
Matthew~J Coak, David~M Jarvis, Hayrullo Hamidov, Andrew~R Wildes, Joseph~AM Paddison, Cheng Liu, Charles~RS Haines, Ngoc~T Dang, Sergey~E Kichanov, Boris~N Savenko, et~al.
\newblock Emergent magnetic phases in pressure-tuned van der waals antiferromagnet feps 3.
\newblock {\em Physical Review X}, 11(1):011024, 2021.

\bibitem{ni2021imaging}
Zhuoliang Ni, AV~Haglund, H~Wang, B~Xu, C~Bernhard, DG~Mandrus, X~Qian, EJ~Mele, CL~Kane, and Liang Wu.
\newblock Imaging the n{\'e}el vector switching in the monolayer antiferromagnet mnpse3 with strain-controlled ising order.
\newblock {\em Nature nanotechnology}, 16(7):782--787, 2021.

\bibitem{zhang2021strain}
Jian-min Zhang, Yao-zhuang Nie, Xi-guang Wang, Qing-lin Xia, and Guang-hua Guo.
\newblock Strain modulation of magnetic properties of monolayer and bilayer feps3 antiferromagnet.
\newblock {\em Journal of Magnetism and Magnetic Materials}, 525:167687, 2021.

\bibitem{fang2023pressure}
Sixue Fang, Quanjun Li, Zonglun Li, Qing Dong, Xiaoling Jing, Chenyi Li, Haiyan Li, Bo~Liu, Ran Liu, and Bingbing Liu.
\newblock Pressure-induced enhancement and retainability of optoelectronic properties of nips3.
\newblock {\em Materials Research Letters}, 11(2):134--142, 2023.

\bibitem{han2021high}
Xu~Han, Pengbo Song, Jie Xing, Zhong Chen, Danyang Li, Guangyuan Xu, Xiaojun Zhao, Fangyuan Ma, Dongke Rong, Youguo Shi, et~al.
\newblock High-performance phototransistors based on mnpse3 and its hybrid structures with au nanoparticles.
\newblock {\em ACS Applied Materials \& Interfaces}, 13(2):2836--2844, 2021.

\bibitem{li2023proximity}
Xiangzhi Li, Andrew~C Jones, Junho Choi, Huan Zhao, Vigneshwaran Chandrasekaran, Michael~T Pettes, Andrei Piryatinski, M{\"a}rta~A Tschudin, Patrick Reiser, David~A Broadway, et~al.
\newblock Proximity-induced chiral quantum light generation in strain-engineered wse2/nips3 heterostructures.
\newblock {\em Nature Materials}, 22(11):1311--1316, 2023.

\bibitem{zhong2023intercalation}
Longsheng Zhong, Hongneng Chen, Wenhu Xie, Weifeng Jia, Yanhe Xiao, Baochang Cheng, Liangxu Lin, and Shuijin Lei.
\newblock Intercalation and defect engineering of layered mnps3 for greatly enhanced capacity and stability in sodium-ion batteries.
\newblock {\em Chemical Engineering Journal}, page 148370, 2023.

\bibitem{mushtaq2023agscp2s6}
Aamir Mushtaq, Mohamed~Yaseen Noor, Ryan Siebenaller, Emma DeAngelis, Adam Fisher, Liam Clink, Justin Twardowski, Gulsum~Kilic Salman, Roberto~C Myers, Emmanuel Rowe, et~al.
\newblock Agscp2s6 van der waals layered crystal: A material with a unique combination of extreme nonlinear optical properties.
\newblock {\em The Journal of Physical Chemistry Letters}, 14(14):3527--3534, 2023.

\bibitem{susner2017metal}
Michael~A Susner, Marius Chyasnavichyus, Michael~A McGuire, Panchapakesan Ganesh, and Petro Maksymovych.
\newblock Metal thio-and selenophosphates as multifunctional van der waals layered materials.
\newblock {\em Advanced Materials}, 29(38):1602852, 2017.

\bibitem{ghobadi2013band}
Nader Ghobadi.
\newblock Band gap determination using absorption spectrum fitting procedure.
\newblock {\em International nano letters}, 3(1):2, 2013.

\bibitem{senkine1989raman}
T~Senkine, M~Jouanne, C~Julien, and M~Balkanski.
\newblock Raman scattering in the antiferromagnet feps3 intercalated with lithium.
\newblock {\em Materials Science and Engineering: B}, 3(1-2):91--95, 1989.

\bibitem{balkanski1987effects}
M~Balkanski, M~Jouanne, G~Ouvrard, and M~Scagliotti.
\newblock Effects due to spin ordering in layered mpx3 compounds revealed by inelastic light scattering.
\newblock {\em Journal of Physics C: Solid State Physics}, 20(27):4397, 1987.

\bibitem{scagliotti1987raman}
M~Scagliotti, M~Jouanne, M~Balkanski, G~Ouvrard, and G~Benedek.
\newblock Raman scattering in antiferromagnetic feps 3 and fepse 3 crystals.
\newblock {\em Physical Review B}, 35(13):7097, 1987.

\bibitem{gjikaj2006rb2p2s6}
Mimoza Gjikaj, Claus Ehrhardt, and Wolfgang Brockner.
\newblock Rb2p2s6--a new alkali thiophosphate: Crystal structure and vibrational spectra of rubidium hexathiodiphosphate (v).
\newblock {\em Zeitschrift f{\"u}r Naturforschung B}, 61(9):1049--1053, 2006.

\bibitem{oliveira20232d}
Filipa~M Oliveira, Jan Pa{\v{s}}tika, Iva Plutnarov{\'a}, Vlastimil Maz{\'a}nek, Karol Struty{\'n}ski, Manuel Melle-Franco, Zden{\v{e}}k Sofer, and Rui Gusm{\~a}o.
\newblock 2d layered bimetallic phosphorous trisulfides mimiiip2s6 (mi= cu, ag; miii= sc, v, cr, in) for electrochemical energy conversion.
\newblock {\em Small Methods}, 7(2):2201358, 2023.

\bibitem{gusmao2017role}
Rui Gusmao, Zdenek Sofer, David Sedmidubsky, Stepan Huber, and Martin Pumera.
\newblock The role of the metal element in layered metal phosphorus triselenides upon their electrochemical sensing and energy applications.
\newblock {\em ACS Catalysis}, 7(12):8159--8170, 2017.

\bibitem{hangyo1988raman}
M~Hangyo, S~Nakashima, A~Mitsuishi, K~Kurosawa, and S~Saito.
\newblock Raman spectra of mnps3 intercalated with pyridine.
\newblock {\em Solid state communications}, 65(5):419--423, 1988.

\bibitem{long2017isolation}
Gen Long, Ting Zhang, Xiangbin Cai, Jin Hu, Chang-woo Cho, Shuigang Xu, Junying Shen, Zefei Wu, Tianyi Han, Jiangxiazi Lin, et~al.
\newblock Isolation and characterization of few-layer manganese thiophosphite.
\newblock {\em ACS nano}, 11(11):11330--11336, 2017.

\bibitem{silipigni2005x}
L~Silipigni, G~Di~Marco, G~Salvato, and V~Grasso.
\newblock X-ray photoelectron spectroscopy characterization of the layered intercalated compound k2xmn1- xps3.
\newblock {\em Applied surface science}, 252(5):1998--2005, 2005.

\bibitem{beechem2008temperature}
Thomas Beechem and Samuel Graham.
\newblock Temperature and doping dependence of phonon lifetimes and decay pathways in gan.
\newblock {\em Journal of Applied Physics}, 103(9), 2008.

\bibitem{jiang2019defect}
Jie Jiang, Tao Xu, Junpeng Lu, Litao Sun, and Zhenhua Ni.
\newblock Defect engineering in 2d materials: precise manipulation and improved functionalities.
\newblock {\em Research}, 2019.

\end{thebibliography}

\end{document}